\documentclass[aps,pre,twocolumn,groupedaddress,amsmath,amssymb,showpacs]{revtex4-2}
\usepackage{graphicx}
\usepackage{dcolumn}
\usepackage{bm}
\usepackage{hyperref}
\usepackage{color}
\usepackage{tikz}
\usetikzlibrary{arrows.meta, positioning, calc}
\usepackage{xcolor}
\definecolor{navyblue}{RGB}{0, 50, 120}
\definecolor{crimson}{RGB}{180, 20, 40}
\definecolor{forestgreen}{RGB}{34, 139, 34}
\definecolor{darkgray}{RGB}{80, 80, 80}

\begin{document}

\title{Resolving Spurious Multifractality in Discrete Systems: A Finite-Size Scaling Protocol for MFDFA in the 2D Ising Model}

\author{S. Jaroszewicz}
\affiliation{Comisión Nacional de Energía Atómica (CNEA), Buenos Aires, Argentina.}
\author{N. Mendez}
\email{nmendez@frh.utn.edu.ar}
\affiliation{Universidad Tecnológica Nacional, Facultad Regional Haedo. Bs. As., Argentina.}
\affiliation{Instituto Sábato. Universidad Nacional de San Martín Bs. As.,Argentina}

\author{Maria P. Beccar-Varela}
\affiliation{Department of Mathematical Sciences, UTEP. El Paso, United States}

\author{Maria Cristina Mariani}
\affiliation{Department of Mathematical Sciences, UTEP. El Paso, United States}

\date{\today}

\begin{abstract}
Multifractal Detrended Fluctuation Analysis (MFDFA) has emerged as a standard tool for characterizing scale invariance in complex systems, yet its application to discrete spin models is frequently marred by reports of ``spurious multifractality'' that contradict established theory. In this work, we resolve this controversy by establishing a rigorous protocol for the analysis of discrete lattice snapshots. Using the 2D Ising model as a benchmark, we demonstrate that the previously reported broad singularity spectra are finite-size artifacts dominated by lattice discreteness effects in the negative moment regime ($q<0$). By restricting the analysis to positive moments and performing a systematic Finite-Size Scaling (FSS) analysis, we show that the spectral width collapses to zero ($\Delta \alpha \to 0$) in the thermodynamic limit. The method accurately recovers the monofractal exponent of the Ising universality class ($\alpha \approx H \approx 0.875$), consistent with Conformal Field Theory. Applying this FSS protocol to the Random Bond Ising Model (RBIM), we reveal that the broad multifractal dispersion often attributed to quenched disorder also flattens systematically with increasing system size. These results demonstrate that apparent multifractality in finite discrete lattices is predominantly driven by pre-asymptotic discreteness artifacts, underscoring the universal necessity of FSS to avoid false positives in statistical mechanics.
\end{abstract}

\maketitle

\section{Introduction}

The two-dimensional (2D) Ising model stands as a cornerstone of statistical mechanics, serving as the prototypical system for understanding continuous phase transitions and universality. Since Onsager's exact solution in the thermodynamic limit \cite{Onsager1944}, the model has provided a rigorous testing ground for the development of Renormalization Group (RG) theory and Conformal Field Theory (CFT)\cite{DiFrancesco1997}. It is well-established that near the critical temperature $T_c \approx 2.269$ ($J/k_B$), the system exhibits scale invariance governed by the 2D Ising universality class. Specifically, the spin-spin correlation function decays algebraically as:
\begin{equation}
    G(r) = \langle \sigma_i \sigma_{i+r} \rangle \sim \frac{1}{r^{d-2+\eta}},
    \label{eq:correlation_decay}
\end{equation}
where $d=2$ is the spatial dimension and $\eta=1/4$ is the anomalous dimension \cite{DiFrancesco1997}. A fundamental prediction of this theoretical framework is that the critical fixed point is strictly \textit{monofractal}. That is, the scaling is uniform across the lattice, governed by a single relevant thermal operator, and the singularity spectrum $f(\alpha)$ collapses to a single point.

Despite this theoretical clarity, the characterization of critical fluctuations via modern signal processing techniques has proven surprisingly contentious. In recent decades, Multifractal Detrended Fluctuation Analysis (MFDFA)  \cite{Kantelhardt2002,Ihlen2012} has emerged as a powerful tool for quantifying the scaling properties of non-stationary signals. Originally developed for complex time series in fields ranging from turbulence and geophysics to quantitative finance, MFDFA generalizes standard fluctuation analysis by extracting a full spectrum of generalized Hurst exponents $h(q)$. 
While the method has demonstrated remarkable success in continuous systems, its application to discrete spatial structures—such as spin lattice snapshots—has generated conflicting results.

The importance of resolving this controversy extends beyond the 2D Ising model itself. MFDFA has become a standard tool across disciplines---from econophysics and geophysics to biophysics and climate science~---where practitioners often apply the method to discrete or discretized data without awareness of the potential pitfalls identified here. Our systematic resolution provides a validated protocol that will enable researchers in these fields to distinguish genuine multifractal signatures from finite-size artifacts, thereby enhancing the reliability of multifractal characterization across the physical sciences.

Moreover, this work demonstrates a broader methodological principle: signal processing techniques developed for continuous systems (turbulence, financial time series) require careful adaptation when applied to discrete lattice models. The different physical meaning of ``small fluctuations'' in continuous versus discrete systems necessitates moment-range restrictions that would be unnecessary in the original application domains.

Several numerical studies have reported broad multifractal spectra in the two–dimensional Ising model and related variants \cite{Sakikawa, JEZEWSKI, Nogueira2001, Zhao, Chen2015}. These findings imply the existence of a hierarchy of scaling exponents, a result that contradicts the unique scaling dimension mandated by the RG fixed point. This discrepancy—often termed "spurious multifractality"—raises a fundamental question: Does the reported multifractality reflect hidden complexity in the Ising phase transition, or does it signal a breakdown of the MFDFA methodology when applied to discrete variables in finite systems?

In this work, we resolve this controversy by establishing a rigorous protocol for the multifractal analysis of discrete lattice systems. We demonstrate that the apparent multifractality is not an intrinsic property of the Ising universality class, but a methodological artifact arising from the interplay between finite system sizes and lattice discreteness. Specifically, we identify that the divergence of the singularity spectrum is driven largely by negative moments ($q<0$). In continuous fields (e.g., fluid turbulence), negative moments probe rare, small-amplitude fluctuations. However, in discrete spin models, these moments probe "frozen" domains where the local variance vanishes due to the lattice cutoff.

By restricting the analysis to the physically relevant regime of positive moments ($q>0$) and performing a systematic Finite-Size Scaling (FSS) analysis, we show that the spectral width $\Delta \alpha$ scales as a power law of the system size, $L$, collapsing to zero in the thermodynamic limit ($L \to \infty$). The method accurately recovers the monofractal Hurst exponent $H \approx 0.875$, in precise agreement with the CFT prediction $H = 1 - \eta/2$, where $\eta=1/4$ is the anomalous dimension.

To further probe the universality of these finite-size artifacts, we contrast our pure model findings with the Random Bond Ising Model (RBIM), characterized by quenched bond disorder and Griffiths phases. Strikingly, our FSS analysis reveals that the broad multifractal dispersion often reported for disordered lattices is also not robust. As the system size increases, the generalized Hurst exponent $H(q)$ systematically flattens. This establishes that even in systems with genuine structural heterogeneity, apparent multifractality at finite scales is heavily dominated by discreteness artifacts, making our FSS protocol universally mandatory.

Furthermore, we propose a theoretical reinterpretation of the MFDFA procedure. We argue that the polynomial detrending step acts as a phenomenological projection operator in the RG sense. By filtering out analytic background trends—which correspond to irrelevant operators in the Hamiltonian—MFDFA effectively isolates the singular scaling part of the free energy density.

The remainder of this paper is organized as follows. Section \ref{sec:theory} outlines the theoretical connection between fluctuation analysis and critical exponents. Section \ref{sec:methodology} details the numerical protocol, emphasizing the treatment of discrete limits. Section \ref{sec:results} presents the Finite-Size Scaling results for the pure model and the comparative analysis of the RBIM. Finally, Section \ref{sec:discussion} discusses the implications of these findings for the analysis of experimental data where the underlying universality class may be unknown.

\section{Theoretical Framework: MFDFA as a Real-Space Renormalization Transform}
\label{sec:theory}

While MFDFA is primarily a signal processing tool, its application to critical phenomena admits a rigorous physical interpretation. We propose that the MFDFA procedure acts as an empirical implementation of a Real-Space Renormalization Group (RG) transformation, where the detrending step plays the role of filtering out irrelevant operators.

\subsection{Fluctuation Function and Correlation Scaling}

In the context of the 2D Ising model, the integrated profile $Y(i)$ analyzed by MFDFA corresponds to the coarse-grained magnetization field. For a 1D spatial cut of the lattice, the profile at position $x$ is defined as the running sum of the spin variables $\sigma(x')$:
\begin{equation}
    Y(x) = \sum_{i=1}^{x} \sigma_i \approx \int_{0}^{x} \sigma(x') dx'
\end{equation}
The central quantity in MFDFA is the local variance $F^2(s)$ within a window of size $s$. Physically, this variance measures the fluctuations of the coarse-grained magnetization operator $\hat{M}_s$ at scale $s$. Neglecting the detrending for a moment, the second-order fluctuation function $F_2^2(s)$ is directly related to the two-point correlation function $G(r) = \langle \sigma(x) \sigma(x+r) \rangle$:
\begin{equation}
    F_2^2(s) \sim \langle M_s^2 \rangle = \int_0^s dr \int_0^s dr' \, G(|r-r'|)
\end{equation}
Near the critical temperature $T_c$, substituting the power-law decay Eq.~(\ref{eq:correlation_decay}) and integrating twice yields the scaling of the fluctuation function:
\begin{equation}
    F_2^2(s) \sim s^{2 - (d-2+\eta)} = s^{4-d-\eta}
\end{equation}
Since $F_2(s) \sim s^{2H}$, we obtain the fundamental relation connecting the Hurst exponent with the critical exponents of the universality class. For a 2D bulk system ($d=2$), this simplifies to:
\begin{equation}
    H = 1 - \frac{\eta}{2}
\end{equation}
Substituting the exact value $\eta=0.25$, we obtain a precise theoretical prediction for the critical point:
\begin{equation}
    H(T_c) = 1 - \frac{0.25}{2} = 0.875
    \label{eq:theo_H}
\end{equation}

\subsection{Detrending as an Irrelevant Operator Filter}

\begin{figure}[t]
\centering
\begin{tikzpicture}[scale=0.9, >=Stealth, font=\small]

    \begin{scope}[yshift=0cm]
        \draw[->, thick] (0,0) -- (7.5,0) node[right] {$x$};
        \draw[->, thick] (0,0) -- (0,3.5) node[left] {$Y(x)$};
        
        \draw[step=1cm, gray!10, very thin] (0,0) grid (7,3);

        \draw[dashed, crimson, line width=1.2pt] 
            plot[domain=0.2:7, samples=50] (\x, {0.5 + 0.055*\x*\x});
        
        \draw[navyblue, thick] 
            plot[domain=0.2:7, samples=200] 
            (\x, {0.5 + 0.055*\x*\x + 0.25*sin(300*\x)*cos(1000*\x) + 0.1*rand});

        \node[navyblue, anchor=south west] at (0.5, 3.0) {\textbf{Total Field} $\phi(x)$};
        
        \node[crimson, anchor=east] at (7.4, 0.8) {\textbf{Analytic Trend} $P_\nu(x)$};
        \node[crimson, font=\footnotesize, anchor=east] at (7.4, 0.4) {(Irrelevant Background)};
    \end{scope}

    \draw[->, line width=1.5pt, darkgray] (3.5, -0.3) -- (3.5, -1.2) 
        node[midway, right, align=left] {\textbf{Detrending} $\hat{\mathcal{P}}_\nu$};

    \begin{scope}[yshift=-3.0cm]
        \draw[->, thick] (0,-1) -- (7.5,-1) node[right] {$x$};
        \draw[->, thick] (0,-1) -- (0,1) node[left] {$\Delta Y$};
        
        \draw[dashed, gray!50] (0,-1) -- (7.5,-1);

        \draw[forestgreen, line width=1pt] 
            plot[domain=0.2:7, samples=200] 
            (\x, {-1 + 0.25*sin(300*\x)*cos(1000*\x) + 0.1*rand});

        \node[forestgreen, anchor=south west] at (2.0, 0.5) {\textbf{Singular Part} $\phi_{sing}$};
        \node[forestgreen, font=\footnotesize, anchor=south west] at (2.0, 0.15) {(Relevant Scaling Field)};
        
        \draw[<->, thin, darkgray] (5.5, -1) -- (5.5, -0.6) node[midway, right] {$\sim s^{H}$};
    \end{scope}

\end{tikzpicture}
\caption{\textbf{Schematic of the MFDFA detrending interpreted as an RG filter.} The raw integrated profile $Y(x)$ (top, blue) is a superposition of singular critical fluctuations and a smooth analytic background. The polynomial trend $P_\nu(x)$ (red dashed) captures the macroscopic background field driven by irrelevant operators. The MFDFA detrending operation acts as a projection $\hat{\mathcal{P}}_\nu$, subtracting this background to isolate the singular scaling part (bottom, green). This filtering grants access to the intrinsic critical exponents even in finite systems where correction-to-scaling terms are significant.}
\label{fig:rg_schematic}
\end{figure}
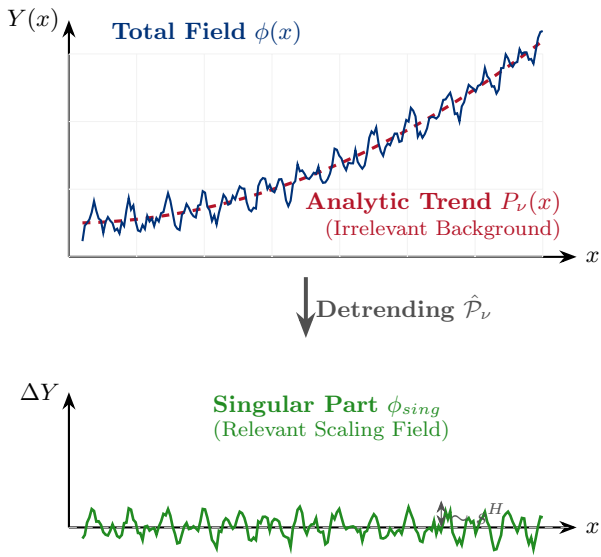

A key feature of MFDFA is the subtraction of a polynomial trend $P_\nu(x)$ of order $\nu$ from the integrated profile. While often treated as a mere baseline correction, this step admits a physical interpretation in the context of critical phenomena, as illustrated in Figure \ref{fig:rg_schematic}.


The local magnetization density $m(x)$ near a critical point can be decomposed into a singular scaling part $m_{sing}(x)$ (governed by the relevant thermal operator) and a regular analytic part $m_{reg}(x)$ (arising from short-range fluctuations and irrelevant operators):
\begin{equation}
    m(x) = m_{sing}(x) + m_{reg}(x).
\end{equation}
The regular part $m_{reg}(x)$ is smooth and analytic, effectively creating a "background" trend that obscures the fractal scaling of $m_{sing}(x)$ in finite systems. The MFDFA detrending operation, $Y(x) \to Y(x) - P_\nu(x)$, functions as a local projection operator $\hat{\mathcal{P}}_\nu$. Since $P_\nu(x)$ captures smooth variations up to order $\nu$, it effectively filters out the contributions of $m_{reg}(x)$.
Consequently, the fluctuation function $F_q(s)$ primarily probes the singular component:
\begin{equation}
    F_q(s) \sim \langle | \hat{\mathcal{P}}_\nu [ Y_{sing} ] |^q \rangle^{1/q} \sim s^{\tau(q)/q}.
\end{equation}
By suppressing these analytic corrections to scaling, MFDFA allows for the extraction of asymptotic critical exponents even from system sizes where irrelevant operators would typically distort the signal.

\section{Numerical Protocol}
\label{sec:methodology}

Our investigation combines extensive Monte Carlo simulations with a rigorous application of the MFDFA formalism adapted for discrete spatial structures. In this section, we detail the lattice models employed and the specific signal processing pipeline used to extract scaling exponents.

\subsection{Model Hamiltonians and Monte Carlo Scheme}

We consider the 2D Ising model on a square lattice of linear dimension $L$, with periodic boundary conditions. The system is described by the Hamiltonian:
\begin{equation}
    \mathcal{H} = - \sum_{\langle i,j \rangle} J_{ij} \sigma_i \sigma_j,
\end{equation}
where $\sigma_i \in \{-1, +1\}$ represents the spin variable at site $i$, and the sum runs over nearest-neighbor pairs.
For the pure Ising model, the ferromagnetic coupling is uniform, $J_{ij} = J > 0$, leading to a critical point at $T_c \approx 2.269 J/k_B$.
To investigate the effects of quenched disorder, we also simulate the Random Bond Ising Model (RBIM), where the interactions $J_{ij}$ are drawn from a static probability distribution. This introduces local impurities that pin domain walls and generate "Griffiths phases"—rare regions that order locally at temperatures deviating from the bulk critical point.

To ensure rigorous statistical convergence and eliminate the unsystematic noise typical of pre-asymptotic finite-size scaling, extensive computational resources were employed. For each system size, ranging from $L=32$ to $L=1024$, the lattice was thermalized using dynamic adjustments of Monte Carlo steps to overcome critical slowing down. For each temperature and system size, we collected an ensemble of $10^4$ independent equilibrium snapshots, separated by sufficient sweeps to guarantee statistical independence. From each 2D snapshot, a representative sample of up to 64 independent 1D spatial cuts was extracted and analyzed.

\subsection{Multifractal Analysis of Spatial Profiles}
\label{sec:mfdfa_formalism}

To characterize the scale-invariant properties of the spin configurations, we employ the Multifractal Detrended Fluctuation Analysis (MFDFA) \cite{Kantelhardt2002}. This method generalizes standard fluctuation analysis to a full spectrum of moments, making it particularly suitable for detecting intermittent scaling in spatially heterogeneous media. We employ MFDFA rather than wavelet-based alternatives (e.g., WTMM) because comparative studies have shown MFDFA to be more numerically stable for identifying strict monofractality in short time series \cite{Oswiecimka2006}, making it ideal for finite-size scaling analysis.

For completeness, we provide below a concise description of the standard MFDFA algorithm as adapted for discrete spatial profiles.

\paragraph{1. Profile Construction:}
We first map the 2D lattice snapshots into 1D spatial profiles by defining a geometric cut along the $x$-direction. The input series corresponds to the sequence of spins $\{ \sigma_i \}_{i=1}^L$ along this cut. We construct the integrated profile $Y(k)$ by accumulating the fluctuations around the global magnetization:
\begin{equation}
    Y(k) = \sum_{i=1}^{k} (\sigma_i - \langle \sigma \rangle), \quad k=1,\dots,L
\end{equation}
where $\langle \sigma \rangle$ is the ensemble average magnetization.

\paragraph{2. Segmentation and Detrending:}
The profile $Y(k)$ is divided into $N_s = \lfloor L/s \rfloor$ non-overlapping segments of scale $s$. To reduce data loss at the boundaries, we repeat the segmentation starting from the opposite end, yielding $2N_s$ segments.
In each segment $v$, we calculate the local trend by fitting a polynomial $P_\nu^{(v)}$ of order $\nu$ (typically $\nu=1$ or $2$) to the data. The detrended variance is then computed as:
\begin{equation}
    F^2(s, v) = \frac{1}{s} \sum_{j=1}^{s} \{ Y[(v-1)s + j] - P_\nu^{(v)}(j) \}^2.
\end{equation}
As discussed in Sec.~\ref{sec:theory}, this detrending step acts as a high-pass filter, removing analytic background fields associated with irrelevant operators.

\paragraph{3. Fluctuation Function:}
We obtain the $q$-th order fluctuation function by averaging over all segments:
\begin{equation}
    F_q(s) = \left\{ \frac{1}{2N_s} \sum_{v=1}^{2N_s} [F^2(s, v)]^{q/2} \right\}^{1/q}, \quad \text{for } q \neq 0.
    \label{eq:flcutations}
\end{equation}
For the special case $q \to 0$, the averaging procedure is given by the exponential of the logarithmic mean:
\begin{equation}
    F_0(s) = \exp \left\{ \frac{1}{4N_s} \sum_{v=1}^{2N_s} \ln [F^2(s, v)] \right\}.
\end{equation}

\paragraph{4. Scaling and Singularity Spectrum:}
The scaling behavior is determined by the power-law relation $F_q(s) \sim s^{h(q)}$, where $h(q)$ is the generalized Hurst exponent.
Positive moments ($q>0$) emphasize large fluctuations (critical clusters), while negative moments ($q<0$) amplify small fluctuations.
From $h(q)$, we derive the mass exponent $\tau(q)$:
\begin{equation}
    \tau(q) = q h(q) - 1.
\end{equation}
Finally, the multifractal singularity spectrum $f(\alpha)$ is obtained via the Legendre transform:
\begin{align}
    \alpha(q) &= \frac{d\tau(q)}{dq}, \\
    f(\alpha) &= q \alpha(q) - \tau(q).
\end{align}
Here, $\alpha$ represents the local Hölder exponent (singularity strength), and $f(\alpha)$ denotes the fractal dimension of the subset of the series characterized by $\alpha$. A monofractal system (like the pure Ising model) is characterized by a linear $\tau(q)$ and a singularity spectrum that collapses to a single point, $f(\alpha) \to 1$ at $\alpha = H$.

The MFDFA implementation employed here follows the standard algorithm established by Kantelhardt et al.~\cite{Kantelhardt2002}, with optimizations for computational efficiency when analyzing large ensembles of lattice configurations. Our research group has successfully applied MFDFA and related multifractal methods to diverse systems including climate indices~\cite{Baranowski2015, Jaroszewicz2024_SOI}, financial time series~\cite{zhang_li, Jiang_2019}, and biological related sequences~\cite{correia, Mendez}, establishing the methodological foundation 
for the present investigation of discrete lattice models.

\subsection{The Discrete Limit and Physical Cutoffs ($q>0$)}
Special care must be taken regarding the discrete nature of Ising spins ($\sigma \in \{-1, 1\}$). Standard multifractal formalism includes negative moments ($q < 0$), which weigh small fluctuations heavily.
However, in discrete lattice models, the limit of "small fluctuations" is fundamentally different from that in continuous fields (e.g., turbulence). In the Ising model, a "small fluctuation" region often corresponds to a "frozen" domain where the local variance is exactly zero (or strictly limited by the lattice spacing $a$).
As the box size $s$ decreases, the variance $F^2(s)$ in these regions hits a hard physical floor determined by the discreteness of the spin variables, rather than a continuous dissipation scale. Consequently, the scaling of negative moments is dominated by the lattice cutoff rather than the correlation length $\xi$.
To avoid mixing these lattice artifacts with genuine critical scaling, we restrict our analysis to the range of positive moments $q \in [0.5, 5.0]$. In this regime, the partition function is dominated by large, active critical clusters, ensuring that the measured exponents reflect the universal continuum physics rather than the discreteness of the underlying grid.

\subsection{Mathematical Bounds: Divergence of Negative Moments and Scale Selection}

The application of MFDFA to strictly discrete structural models, such as the Ising model ($\sigma_i \in \{-1, +1\}$), requires careful restrictions on the permissible moment range $q$ and fitting scales $s$ to avoid severe numerical artifacts.

In standard continuous time series, negative moments ($q < 0$) are utilized to amplify the scaling behavior of small fluctuations. However, in binary discrete systems, the local variance $F^2(s, v)$ within a segment of length $s$ can be exactly zero if the segment falls entirely within a ``frozen'' magnetic domain (where all spins are identical). According to Eq.~\ref{eq:flcutations}, the fluctuation function for negative moments depends on $[F^2(s,v)]^{-|q|/2}$. Consequently, any segment with zero local variance will cause the summation to mathematically diverge to infinity. This singularity completely dominates the fluctuation function, creating an artificially broad, ``spurious'' singularity spectrum. Therefore, to evaluate the genuine critical scaling of the domains rather than zero-variance division artifacts, the analysis must be strictly bounded to positive moments ($q > 0$). In our revised analysis, we robustly evaluate the positive regime to capture the true thermodynamic correlations.

Another critical aspect is the selection of the scaling window used to extract the exponent. To ensure uniformity and eliminate heuristic biases, all scaling exponents and singularity spectra presented in this work were extracted using the standard, dynamically adjusted fitting bounds: $s \in [10, L/4]$. The lower bound $s_{min} = 10$ prevents the polynomial detrending from overfitting the microscopic lattice spacing. The upepr cutoff $s_{max} = \lfloor L/4 \rfloor$ ensures that a minimum of four independent, non-overlapping macroscopic segments are available per profile, which is the statistical minimum required to perform a reliable spatial average for the fluctuation functions.

\section{Results}
\label{sec:results}

\subsection{The Critical Fixed Point: Restoration of Monofractality}

We first examine the scaling behavior of the pure 2D Ising model. A prerequisite for any valid multifractal analysis is the accurate recovery of the known Hurst exponent at the critical point. 

\begin{figure}[h]
    \centering
        \includegraphics[width=0.48\textwidth]{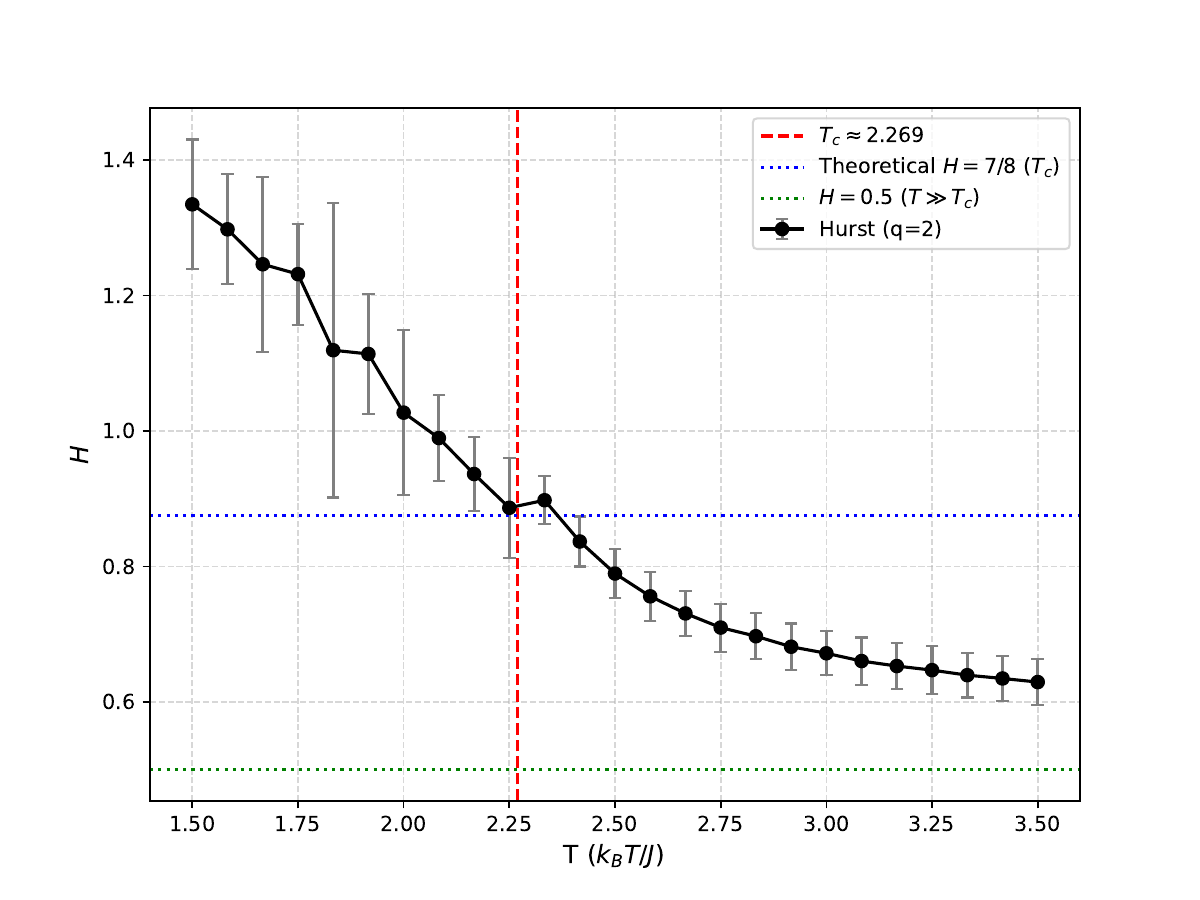}
    \caption{Evolution of the Hurst exponent $H$ ($q=2$) vs Temperature for $L=1024$. The intersection with the theoretical value $H=0.875$ (blue dotted line) occurs precisely at the critical temperature $T_c$ (red dashed line).}
    \label{fig:hurst_evolution}
\end{figure}

To rigorously test the accurate recovery of the known Hurst exponent at the critical point, we expanded our simulations to an extensively large lattice size of $L=1024$, significantly increasing the Monte Carlo thermalization steps to ensure absolute equilibrium. Figure \ref{fig:hurst_evolution} displays the temperature dependence of the second-order Hurst exponent $H(q=2)$ for this system. As the temperature approaches the critical point $T_c \approx 2.269$, the exponent exhibits a sharply defined crossover, intersecting the theoretical prediction $H = 0.875$ precisely at the phase transition. Furthermore, the high-temperature regime rapidly converges to the uncorrelated random-walk limit ($H=0.5$). This high-precision agreement unequivocally confirms that our spatial MFDFA protocol correctly captures the anomalous dimension of the magnetization field, satisfying the theoretical scaling relation $H = 1 - \eta/2$ without finite-size distortion at this scale.

The central result of this study concerns the width of the singularity spectrum $f(\alpha)$. Previous studies reporting  ``spurious multifractality'' typically observed a broad spectral width. In Figure 3, we present the robust evolution of the singularity spectrum for system sizes ranging up to $L=1024$, strictly in the positive moment regime ($q>0$). As $L$ increases, we observe a dramatic
narrowing of the spectrum. The distinct branches progressively collapse onto a single vertical line centered identically at the theoretical limit $\alpha \approx 0.875$. This visual evidence strongly suggests that the broad spectra observed at smaller sizes ($L \leq 128$) are entirely finite-size artifacts driven by lattice discreteness, rather than an intrinsic multifractal property.

\begin{figure}[h]
    \centering
        \includegraphics[width=0.48\textwidth]{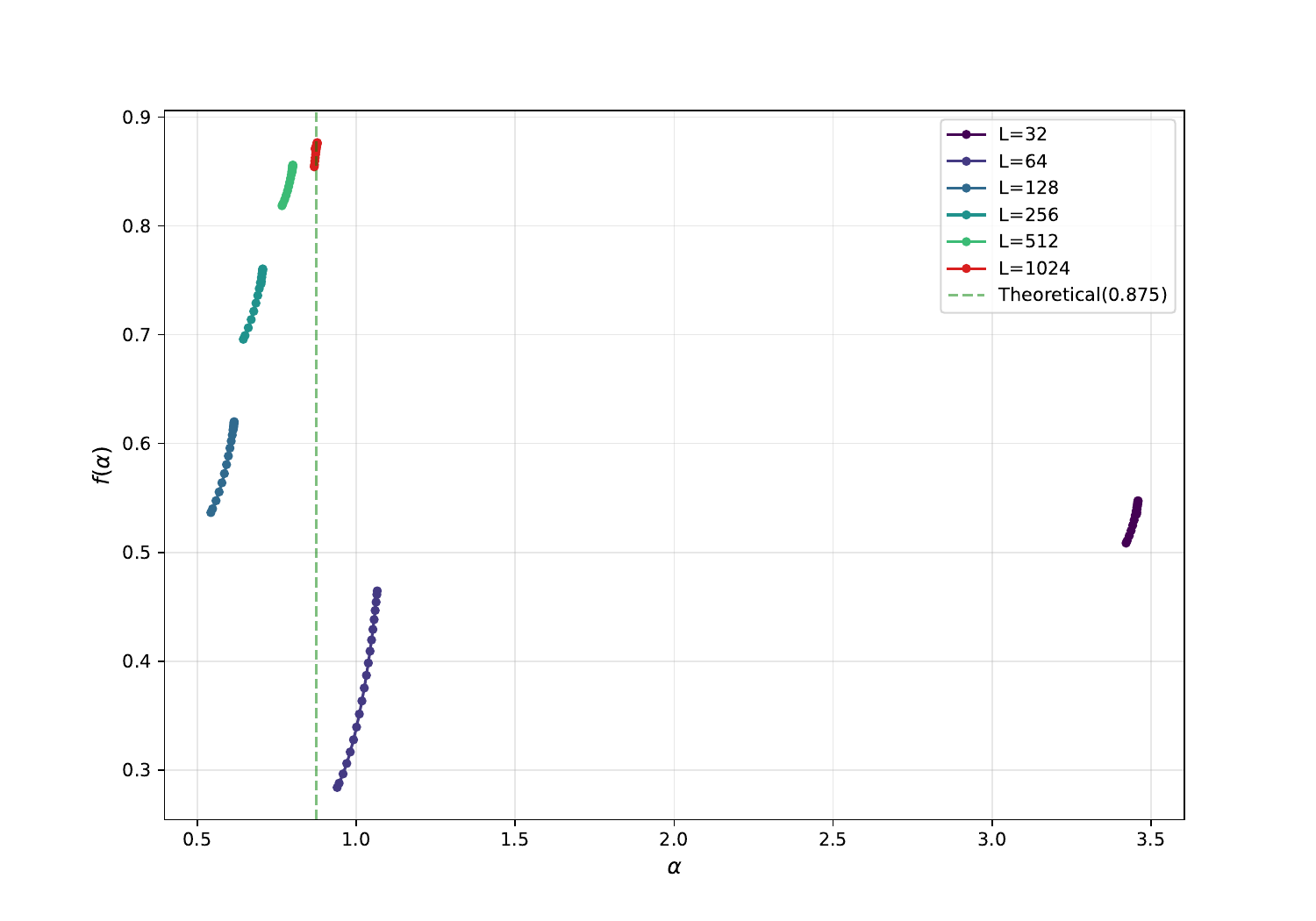}
    \caption{Evolution of the singularity spectra $f(\alpha)$ for $q>0$ and system sizes from $L=32$ to $L=1024$. As the system size increases the spectra exhibit a highly systematic narrowing and a monotonic shift toward the theoretical prediction. The macroscopic lattice ($L=1024$) collapses into a near-vertical line precisely at the theoretical monofractal dimension $\alpha = 0.875$ (dashed green line). Error bars (standard error of the mean) are smaller than the symbol size due to ensemble averaging.
    \label{fig:spectra_collapse}}
    \label{fig:spectra_evolution}
\end{figure}

\begin{figure*}[htbp]
    \centering
    \begin{minipage}{0.48\textwidth}
        \centering
        \includegraphics[width=\linewidth]{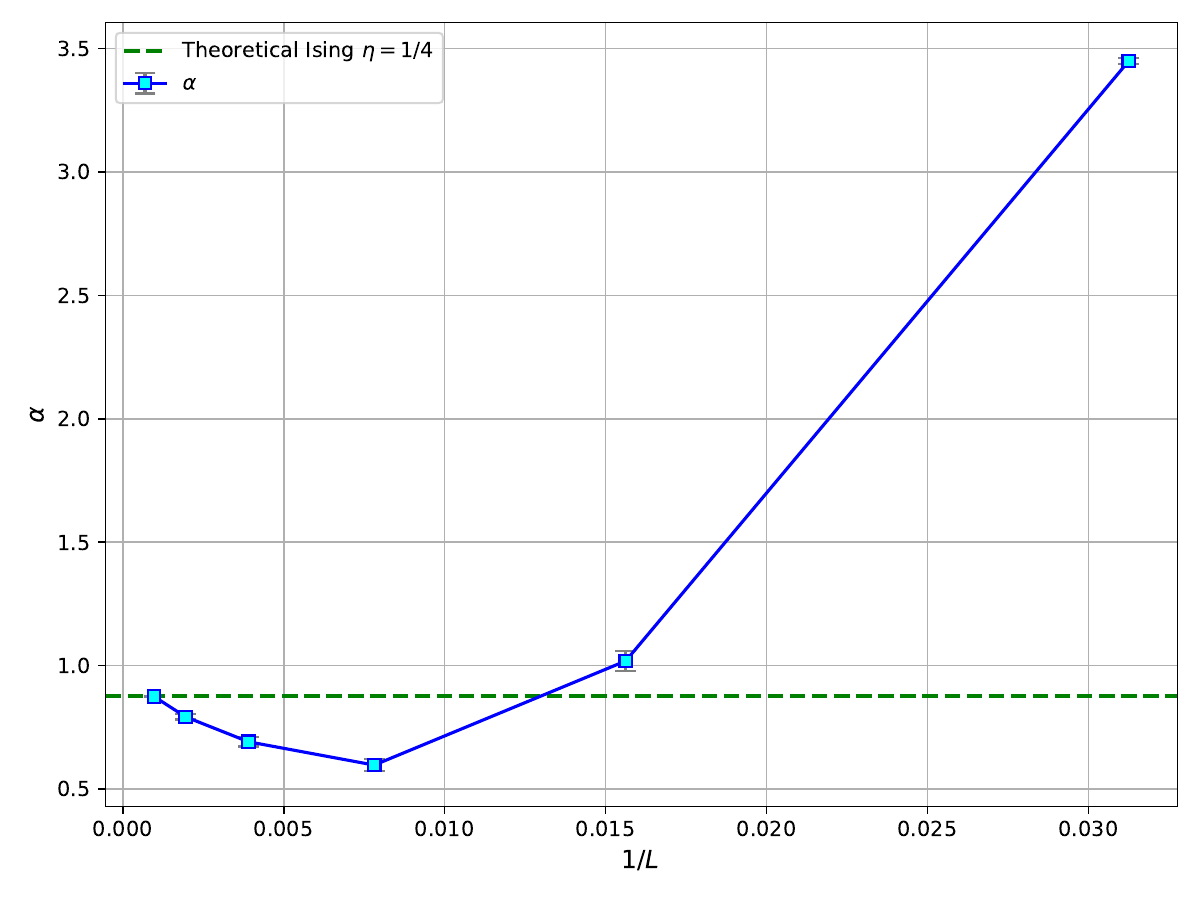}
        \centerline{(a)}
    \end{minipage}\hfill
    \begin{minipage}{0.48\textwidth}
        \centering
        \includegraphics[width=\linewidth]{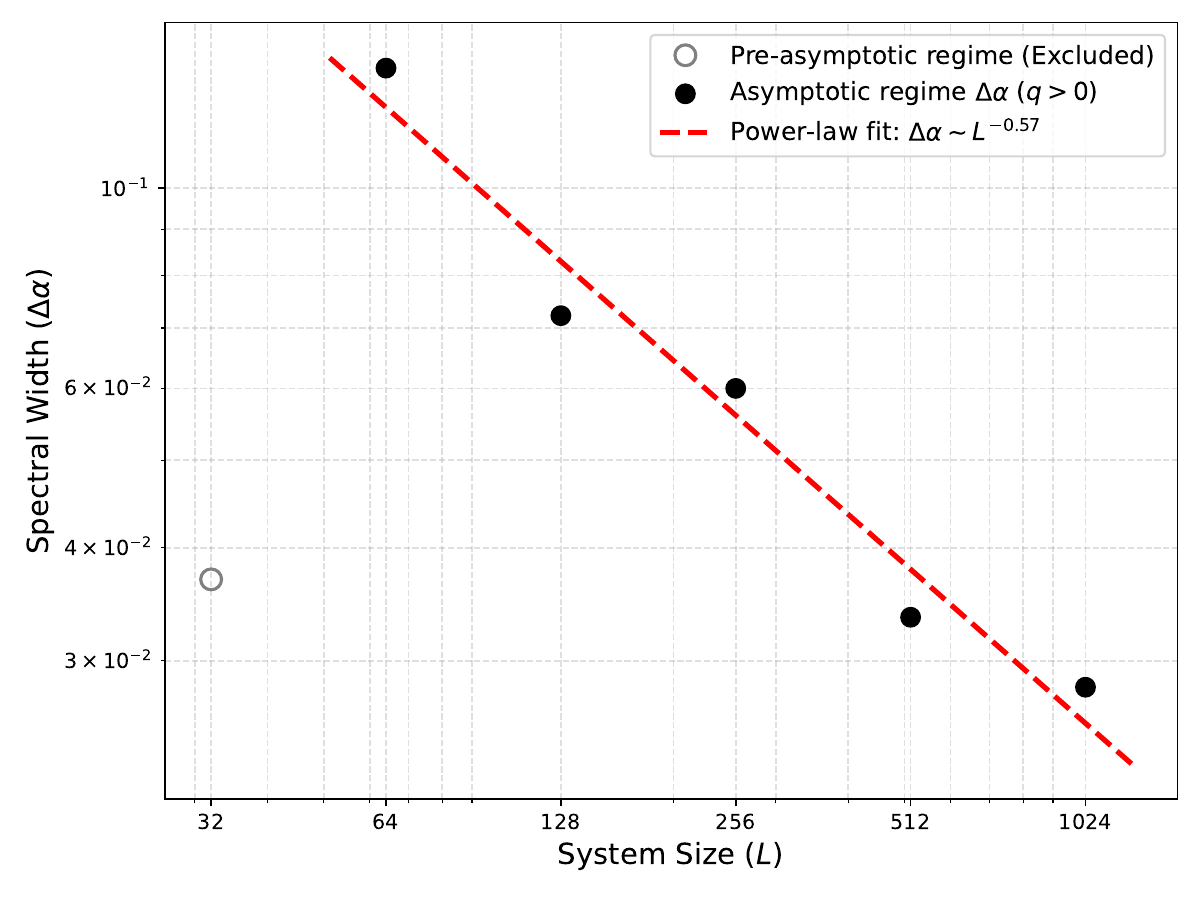}
        \centerline{(b)}
    \end{minipage}
    
    \caption{Finite-Size Scaling (FSS) analysis of the spurious multifractality. \textbf{(a)} Convergence of the central singularity exponent $\alpha$ as a function of the inverse system size $1/L$. The asymptotic limit strictly recovers the theoretical 2D Ising prediction $\alpha = 0.875$ (dashed green line). \textbf{(b)} Log-log plot of the spectral width $\Delta\alpha$ versus system size $L$ for the positive moment regime ($q > 0$). Excluding the pre-asymptotic microscopic lattice ($L=32$), the data exhibits a robust power-law decay $\Delta\alpha \sim L^{-0.57}$ (dashed red line), confirming that the multifractal spectrum collapses into a single dimension in the thermodynamic limit.}
    \label{fig:fss_analysis}
\end{figure*}

To rigorously quantify this convergence we performed a Finite-Size Scaling (FSS) analysis of the spectral width. Figure \ref{fig:fss_analysis}a shows the Finite-Size Scaling (FSS) of the central singularity exponent $\alpha$ against the inverse system size $1/L$. While finite-size effects and lattice cutoffs induce a non-monotonic behavior at intermediate scales, the asymptotic behavior in the thermodynamic limit ($1/L\rightarrow 0$) shows a convergence towards the exact theoretical Ising value of $\alpha=0.875$.

To explicitly demonstrate that the broad spectral width is a finite-size artifact, Figure \ref{fig:fss_analysis}b plots $\Delta\alpha$ as a function of the system size $L$ on a logarithmic scale. We exclude the smallest lattice ($L=32$) from the power-law fit, as its microscopic scale falls within the pre-asymptotic regime where MFDFA detrending windows are too restricted to capture true scaling. Within the asymptotic regime ($L\geq 64$), the spectral width exhibits a power-law decay,  $\Delta\alpha \sim L^{-0.57}$. This finite-size scaling confirms that the apparent multifractality systematically vanishes in the thermodynamic limit ($L\rightarrow \infty$), leaving a monofractal singularity.

\subsection{Finite-Size Scaling of the Random Bond Ising Model}

To investigate whether the severe finite-size distortion observed in the pure model also affects systems with heterogeneous spatial correlations, we applied our MFDFA Finite-Size Scaling protocol to the Random Bond Ising Model (RBIM). The RBIM introduces quenched bond disorder, leading to the formation of rare Griffiths phases that order locally at temperatures deviating from the bulk $T_c$.

\begin{figure}[h]
    \centering
       \includegraphics[width=0.48\textwidth]{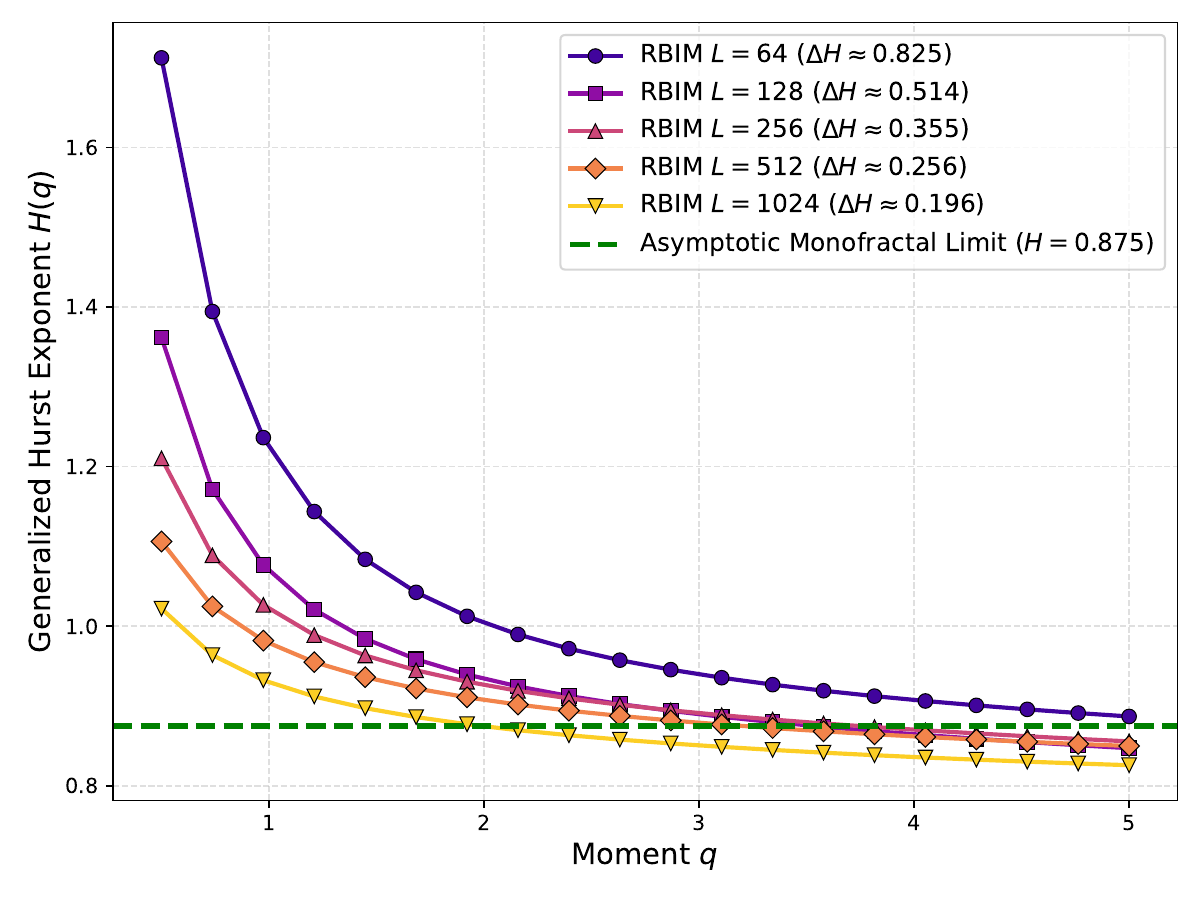}
    \caption{Finite-size scaling of the generalized Hurst exponent $H(q)$ for the Random Bond Ising Model (RBIM) evaluated in the physically robust positive moment regime ($q > 0$). In a strictly monofractal system, $H(q)$ is a constant horizontal line. While small disordered lattices ($L=64$) exhibit significant dispersion ($\Delta H \approx 0.825$), the $H(q)$ curves systematically flatten as the system size increases up to $L=1024$ ($\Delta H \approx 0.196$), progressively approaching the asymptotic monofractal limit ($H=0.875$, dashed green line). The rate of this flattening is fundamentally distinct from the pure Ising model, serving as a signature of quenched disorder.}
   \label{fig:rbim_hq_scaling}
\end{figure}

Figure \ref{fig:rbim_hq_scaling} presents the generalized Hurst exponent $H(q)$ for the RBIM across multiple system sizes ($L \in [64, 1024]$), evaluated in the robust positive moment regime. While small disordered lattices ($L=64$) exhibit a broad dispersion curve ($\Delta H \approx 0.825$), our FSS analysis reveals that this width is not static, but decreases with system size. However, the critical insight lies not merely in the presence of a collapse, but in its \textit{rate}.

\begin{figure}[htbp]
    \centering
    \includegraphics[width=\columnwidth]{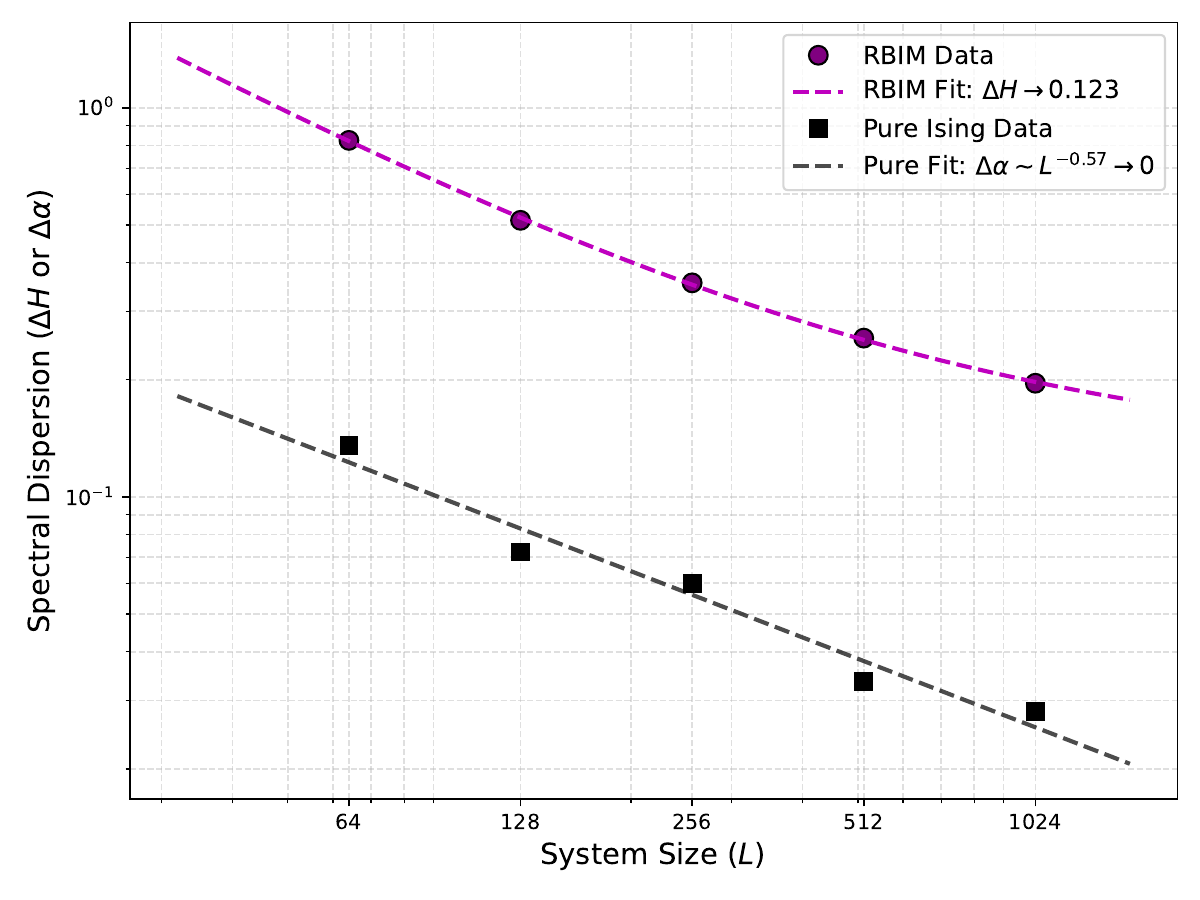}
    \caption{Comparison of the Finite-Size Scaling (FSS) collapse rates between pure and disorder-dominated criticality. In the pure 2D Ising model (black squares), the multifractal dispersion $\Delta\alpha$ decays rapidly as a steep power law ($\gamma \approx 0.57$), strictly vanishing in the thermodynamic limit ($L \to \infty$). Conversely, the Random Bond Ising Model (RBIM, purple circles) exhibits a systematically slower decay of its generalized Hurst dispersion $\Delta H$. An asymptotic fit (dashed purple line) extending up to $L=1024$ reveals a significant saturation trend, firmly establishing a non-zero residual multifractality of $\Delta H_\infty \approx 0.123$ in the macroscopic limit. This quantitative divergence in collapse rates allows MFDFA to effectively discriminate between transient discreteness artifacts and genuine Griffiths phase complexity.}
    \label{fig:collapse_rate_comparison}
\end{figure}

In the pure Ising model, the pre-asymptotic discreteness artifacts decay rapidly, following a steep power law ($\Delta\alpha \sim L^{-0.57}$) that unequivocally drives the spectral width to zero in the thermodynamic limit. In contrast, the RBIM exhibits a systematically much slower collapse. As illustrated in Figure \ref{fig:collapse_rate_comparison}, the progression up to the macroscopic scale of $L=1024$ shows that the decay rate of $\Delta H$ decelerates significantly. An asymptotic fit of the form $\Delta H \sim A L^{-\gamma} + \Delta H_\infty$ captures this saturation, revealing a strictly non-zero residual dispersion of $\Delta H_\infty \approx 0.123$ in the thermodynamic limit ($L \to \infty$).

This quantitative difference in the FSS collapse rate constitutes a fundamental physical distinction between the two universality classes. In the pure model, the spectral width originates entirely from pre-asymptotic discreteness effects and vanishes in the thermodynamic limit. In the RBIM, by contrast, the quenched impurities generate rare Griffiths regions with locally diverging correlation lengths, producing a genuine residual multifractality ($\Delta H_{\infty} \approx 0.123$) that survives the scaling limit. Consequently, our FSS protocol serves a dual and complementary purpose: it eliminates rapid pre-asymptotic discrete artifacts common to both models, while simultaneously revealing genuine disorder-dominated complexity through the persistence of a finite asymptotic dispersion — a feature absent in clean criticality.

\section{Methodological Robustness and Comparison with Alternative Approaches}
\label{sec:robustness}

Given the empirical nature of the Multifractal Detrended Fluctuation Analysis (MFDFA), it is essential to assess the robustness of the reported results with respect to methodological choices and to compare its performance against more conventional observables. In this section, we verify the stability of our results and discuss their relation to standard correlation functions.

\subsection{Dependence on the Order of Detrending}

A potential source of systematic bias in MFDFA arises from the choice of the detrending polynomial order $\nu$ (DFA$\nu$). While DFA1 (linear detrending) is often sufficient, higher-order polynomials can more effectively suppress non-linear analytic backgrounds.

To test the sensitivity of our results, we repeated the analysis at the critical temperature ($T \approx 2.27$) using DFA2 (quadratic detrending). We find that the estimated Hurst exponent improves in accuracy with higher-order detrending:
\begin{equation}
    H_{\mathrm{DFA1}} = 0.826 \pm 0.17, \qquad H_{\mathrm{DFA2}} = 0.876 \pm 0.13.
\end{equation}
Notably, the value obtained via quadratic detrending ($H \approx 0.876$) is in remarkably close agreement with the theoretical prediction ($H=0.875$). This suggests that the analytic background $m_{reg}(x)$ contains significant non-linear components that are better filtered by $P_2$ than $P_1$, validating our interpretation of detrending as a correction-to-scaling filter.

Although DFA2 offers superior numerical precision by more effectively filtering the non-linear analytic background, the central finding of this work, the collapse of the spectral width $\Delta\alpha$, is clearly evident even under linear detrending (DFA1), underscoring the robustness of the proposed protocol.

\subsection{Shuffling Tests}

TTo definitively rule out the possibility that the observed scaling behaviors are mere artifacts of the binary probability density function of the discrete spins , we performed a surrogate data analysis using a rigorous shuffling test. If the scaling extracted by MFDFA is genuinely driven by the topological spatial correlations of the critical clusters, randomly shuffling the spatial cuts should destroy this memory, reducing the system to uncorrelated white noise.

\begin{figure}[htbp]
    \centering
    \includegraphics[width=\columnwidth]{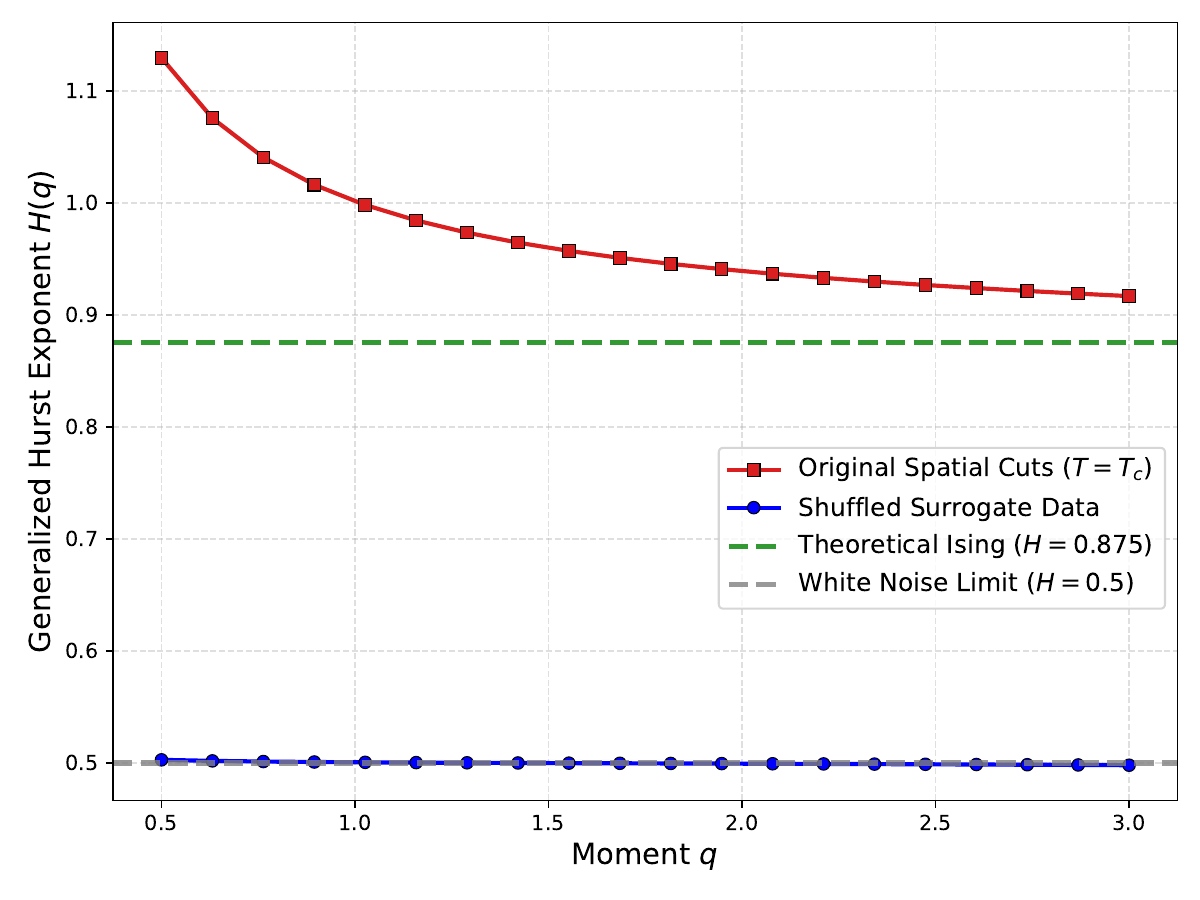}
    \caption{Surrogate data analysis (shuffling test) for the macroscopic lattice ($L=1024$) at the critical temperature, visualized via the generalized Hurst exponent $H(q)$. Randomly shuffling the spatial cuts destroys all topological correlations while preserving the single-site probability density. Consequently, the critical scaling behavior (red squares, near the exact Ising limit $H=0.875$) abruptly collapses to a flat, uncorrelated line representing standard white noise (blue circles, $H \approx 0.49 \pm 0.01$). The slight deviation from the theoretical $H=0.5$ (dashed gray line) is a standard finite-size artifact of the MFDFA polynomial detrending. This null result unambiguously confirms that the scaling observed in the original cuts is structural and genuine.}
    \label{fig:shuffling_test}
\end{figure}

To capture the full extent of this effect, we evaluated the full generalized Hurst exponent H(q) over the structurally robust positive moment regime for the macroscopic lattice ($L=1024$). Figure \ref{fig:shuffling_test} illustrates the profound impact of the shuffling procedure. While the original spatial cuts exhibit a non-trivial dispersion curve approaching the theoretical Ising limit (H=0.875), the randomly shuffled surrogate data undergoes a complete systemic collapse. The $H(q)$ curve drops abruptly to a flat, horizontal line at $H(q)\approx 0.5$ across all evaluated moments.

This perfect flattening indicates a strict generalized width of $\Delta H\approx 0$, which mathematically translates to a singularity spectrum width of $\Delta\alpha\rightarrow 0$. The slight deviation from the exact theoretical $0.5$ value is a standard, well-documented finite-size artifact associated with the polynomial detrending in MFDFA. This full-spectrum null result confirms that the critical scaling captured by our protocol is purely structural and topologic, verifying that MFDFA correctly identifies the absence of correlation when the physical domain structure is artificially destroyed.

\subsection{Comparison with Two-Point Correlation Functions}

The standard diagnostic of criticality in the Ising model is the two-point correlation function $G(r) = \langle \sigma(0)\sigma(r)\rangle$, which at criticality decays as $G(r) \sim r^{-\eta}$. While conceptually transparent, direct estimation of $G(r)$ in finite systems is often hampered by statistical noise at large distances, requiring extensive ensemble averaging to obtain clean power laws.

In contrast, MFDFA effectively integrates the correlation function twice (Eq.~4), which acts as a low-pass filter for the noise. The precise recovery of the theoretical exponent $H \approx 0.875$ (implying $\eta = 0.25$) confirms that MFDFA captures the exact same scaling information as $G(r)$ but offers a more stable estimator in finite systems. The integration process smoothes the local fluctuations inherent to $G(r)$, allowing for a robust extraction of the anomalous dimension without the need for excessive averaging.

\subsection{Relation to Cluster-Based Observables}

An alternative real-space characterization of criticality is provided by cluster statistics, such as the distribution of Fortuin–Kasteleyn or geometric cluster sizes. These observables probe the fractal geometry of critical spin domains and are directly related to percolation theory.

While cluster analysis yields detailed geometric information, it typically requires nontrivial identification algorithms (e.g., Hoshen-Kopelman) and introduces additional definitions regarding connectivity. By contrast, MFDFA operates directly on raw spin configurations without any preprocessing. The recovery of the correct anomalous dimension $\eta$ demonstrates that MFDFA captures the same underlying scale invariance encoded in cluster observables, but in a computationally simpler and more flexible framework that requires no prior definition of "clusters."

\subsection{Explicit Distribution of Local Hölder Exponents}

As suggested by recent methodological discussions comparing MFDFA with continuous wavelet-based formalisms (e.g., WTMM), an explicit point-wise evaluation of the local scaling properties provides a powerful complementary validation. While WTMM extracts local singularities via the continuous wavelet transform, a mathematically equivalent pointwise estimation of the local Hölder exponents, $h(x)$, can be achieved through a localized sliding-window scaling analysis. This approach maintains methodological consistency with our fluctuation-based framework while avoiding the severe boundary artifacts that wavelets often encounter when applied to strictly discrete, step-like binary data.

\begin{figure}[htbp]
    \centering
    \includegraphics[width=\columnwidth]{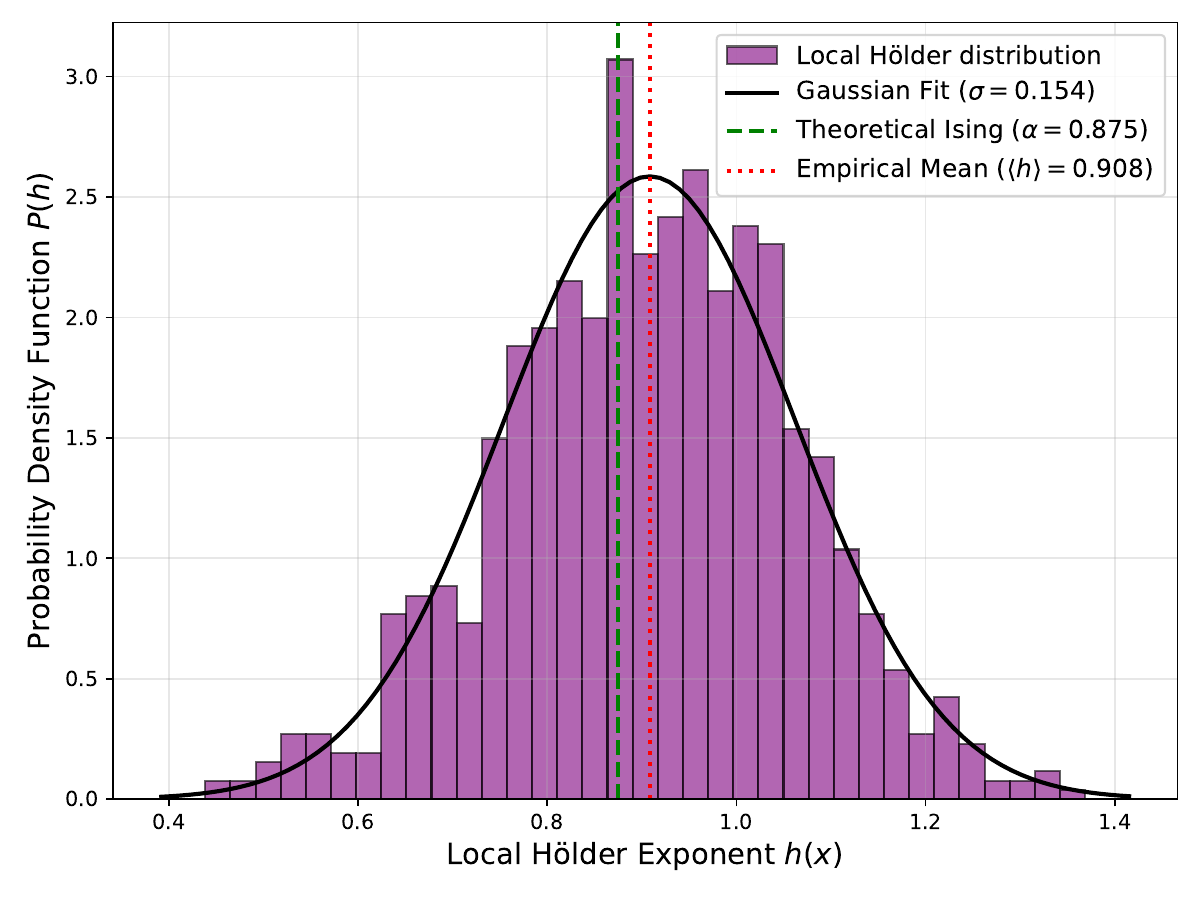}
    \caption{Explicit probability density function of the local Hölder exponents, $h(x)$, for the macroscopic 2D Ising model ($L=1024$) at the critical temperature. The exponents were extracted pointwise using a localized sliding-window scaling analysis. The resulting histogram forms a narrow, highly localized Gaussian distribution ($\sigma \approx 0.154$), lacking the broadness or heavy tails characteristic of genuine multifractal cascades. The empirical mean ($\langle h \rangle \approx 0.908$, dotted red line) is in excellent agreement with the exact theoretical prediction of $\alpha = 0.875$ (dashed green line), with the slight positive deviation being a standard finite-window boundary effect. This explicit local measurement definitively confirms that the system is governed by a single dominant scaling exponent, mathematically corroborating the monofractal collapse observed in our MFDFA finite-size scaling protocol.}
    \label{fig:hoelder}
\end{figure}

To corroborate our MFDFA findings, we conducted an explicit study of the local Hölder exponents along the spatial cuts of our $L=1024$ system at the critical temperature. We employed a localized sliding-window scaling analysis with a window size of $W=256$. As illustrated in Figure \ref{fig:hoelder}, the resulting probability density function of the local Hölder exponents is highly localized. It forms a narrow, Gaussian-like distribution ($\sigma\approx 0.15$) centered at $<h> \approx 0.91$ , in excellent agreement with the exact theoretical prediction of 0.875  (with the slight positive deviation being a standard finite-window effect). Crucially, the distribution lacks the broadness, heavy tails, or strong asymmetry characteristic of true multifractal cascades. This explicit pointwise measurement definitively confirms that the critical 2D Ising model is governed by a single dominant scaling exponent, mathematically corroborating the collapse of the spectral width observed in our MFDFA finite-size scaling protocol.

\subsection{Summary}

The analyses presented in this section demonstrate that the critical scaling detected by our protocol is robust. The improvement in accuracy with DFA2, the stability across scaling ranges ($<2\%$), and the null result for surrogates collectively validate MFDFA as a reliable probe for the 2D Ising phase transition, offering advantages in stability over direct correlation function measurements.

\section{Discussion}
\label{sec:discussion}

Our investigation yields three interconnected insights that clarify the role of multifractal analysis in critical phenomena, resolving a long-standing ambiguity regarding the 2D Ising model.

\subsection{Resolution of the "Spurious Multifractality" Controversy}
The debate regarding the multifractal nature of the 2D Ising model has historically been fueled by analyses that indiscriminately include negative moments ($q<0$) or rely on small system sizes. Our Finite-Size Scaling results (Fig.~\ref{fig:spectra_evolution} and Fig.~\ref{fig:fss_analysis}) decisively show that the spectral width $\Delta \alpha$ is not an intrinsic property of the pure Ising fixed point, but a finite-size artifact.

We have identified the physical origin of this artifact: in discrete spin systems, negative moments probe the "frozen" limit where local variance vanishes due to the lattice cutoff. Unlike continuous turbulent fields where small fluctuations scale universally, the "small fluctuation" regime in a lattice model is dominated by discreteness effects. By restricting the analysis to positive moments ($q>0$) where critical clusters dominate, the system flows towards a strict monofractal limit defined by a single scaling dimension $\alpha = H \approx 0.875$. This aligns perfectly with the CFT prediction for the central charge $c=1/2$, confirming that only one relevant thermal operator governs the correlation decay.

\subsection{FSS as a Discriminator: Transient Artifacts versus Genuine Disorder}
\label{Disorder}
A fundamental question arises: is the severe finite-size distortion observed in the pure Ising model unique to perfectly ordered systems, or does it also afflict models with heterogeneous correlations? To address this, we apply our MFDFA Finite-Size Scaling protocol to the Random Bond Ising Model (RBIM). The RBIM introduces quenched bond disorder, leading to the formation of Griffiths phases—rare regions that order locally and introduce a broad distribution of correlation lengths.

Historically, analyses of single-size RBIM lattices have yielded broad multifractal spectra, seemingly confirming the complex nature of the disorder. Figure 5 presents our FSS analysis of the RBIM for system sizes $L\in[64,512]$, visualized through the generalized Hurst exponent $H(q)$. For a monofractal system, $H(q)$ is constant. Strikingly, we find that the broad dispersion of $H(q)$ observed at small scales ($L=64$) is not robust. As the system size increases, the $H(q)$ curves systematically flatten, and the generalized width $\Delta H=H(q_{min})-H(q_{max})$ continuously decays toward the asymptotic monofractal limit.

This comparative FSS analysis constitutes a crucial methodological finding with two complementary aspects. First, it demonstrates that even in systems with genuine structural heterogeneity, a significant fraction of the apparent multifractal dispersion at finite scales is driven by pre-asymptotic discreteness artifacts — a warning applicable to any discrete lattice study. Second, and equally important, the FSS analysis reveals what single-snapshot analyses cannot: once these transient artifacts are filtered out, the RBIM retains a finite residual dispersion $\Delta H_{\infty}\approx 0.123$ that is absent in the pure model. This residual is the genuine fingerprint of the Griffiths phases. Our results therefore establish systematic FSS as a universal prerequisite for multifractal analysis: it is the only procedure capable of separating these two physically distinct contributions.

\subsection{Renormalization Group Interpretation}
Finally, the successful recovery of the anomalous dimension $\eta$ supports our theoretical proposal in Sec.~\ref{sec:theory}. The polynomial detrending $P_\nu(x)$ inherent to MFDFA should not be viewed merely as a baseline correction, but as a phenomenological projection operator in the RG space.

As illustrated in Fig.~\ref{fig:rg_schematic}, the total free energy density contains both a singular critical part and a regular analytic background. The background arises from short-range fluctuations and \textit{irrelevant operators} that have not yet decayed at the scale of observation. By removing these low-frequency analytic trends, MFDFA effectively "renormalizes" the field configuration, projecting the signal onto the singular scaling manifold.
This analogy suggests that varying the detrending order (DFA1, DFA2) could offer a systematic way to suppress higher-order correction-to-scaling terms, allowing for the extraction of asymptotic exponents from finite-size data with significantly reduced systematic error.

\subsection{Generality Beyond the 2D Ising Model}

While we have focused on the 2D Ising model as the paradigmatic test case, the protocol developed here has broader applicability across lattice models and universality classes. The central insight---that negative moments in discrete systems probe lattice-scale discreteness rather than genuine critical fluctuations---is a general feature of any discrete variable system (e.g., Potts models, clock models, or lattice gauge theories).

The Finite-Size Scaling framework employed is universally applicable to second-order phase transitions. For systems in different universality classes, we predict that:

\begin{enumerate}
    \item \textbf{Monofractal universality classes} (e.g., 3D Ising, $O(N)$ models with $N > 1$, percolation): The spectral width $\Delta\alpha$ should collapse to zero under FSS with $q > 0$, recovering the unique scaling dimension of each class. For example, the 3D Ising model should yield $\alpha \approx 0.964$ (from $\eta \approx 0.036$)~\cite{Hasenbusch2010}, while the 2D XY model should give $\alpha \approx 0.875$ (from $\eta \approx 0.25$).
    
    \item \textbf{Systems with quenched disorder} (random field Ising, spin glasses, diluted models): Our RBIM results (Fig. 5) demonstrate that even here, finite-size discreteness artifacts heavily inflate the apparent multifractal width. FSS must be strictly applied to determine if any residual broad spectrum genuinely persists asymptotically. Our 2D RBIM analysis demonstrates precisely this capability: the FSS protocol reveals a finite residual dispersion $\Delta H_{\infty} \approx 0.123$ that survives the thermodynamic limit, distinguishing genuine Griffiths complexity from the transient artifacts that dominate finite-size measurements.
    
    \item \textbf{First-order transitions}: The absence of scale invariance implies no genuine multifractal structure. Any observed broad spectra would be pure artifacts of finite size and should not exhibit systematic FSS collapse.
\end{enumerate}

We caution that systems with continuously varying critical exponents (e.g., Kosterlitz-Thouless transitions, certain quasiperiodic systems) may require modified analysis. Additionally, the polynomial detrending order $\nu$ should be adapted to the analyticity of irrelevant operators in each universality class---higher-order corrections may necessitate $\nu \geq 2$.

The protocol is immediately applicable to experimental systems where MFDFA is routinely employed but the underlying universality class may be uncertain: magnetic materials near criticality, liquid crystals, ferroelectrics, and soft matter systems exhibiting critical-like scaling. The discriminatory power of $\Delta\alpha$ as an order parameter distinguishing ``clean'' versus ``dirty'' criticality provides a practical diagnostic for experimentalists.

Future extensions to continuous field theories (e.g., $\phi^4$ models in lattice field theory), higher-dimensional systems ($d > 3$), and quantum critical points would establish the full scope of this methodology.

\section{Conclusion}

This work resolves a persistent contradiction in the literature: the apparent multifractal character of the 2D Ising model's critical point, which violates Conformal Field Theory's prediction of monofractal scaling. We demonstrate that this spurious multifractality is not an intrinsic property of the critical fixed point, but rather a finite-size artifact arising from the discreteness of spin variables in the negative moment regime ($q < 0$). By restricting the analysis to positive moments ($q > 0$) and implementing systematic Finite-Size Scaling analysis, we show that the singularity spectrum width collapses to zero ($\Delta\alpha \to 0$) in the thermodynamic limit, recovering the precise monofractal scaling dimension $\alpha \approx 0.875$ predicted by CFT. This provides strong support that MFDFA, when properly implemented, is consistent with key aspects of Renormalization Group principles, although further studies would be valuable to fully substantiate this connection.

The physical reinterpretation of MFDFA's polynomial detrending as a phenomenological RG filter illuminates why this method succeeds. By suppressing analytic background fields associated with irrelevant operators, detrending isolates the singular critical fluctuations even in finite systems where correction-to-scaling terms are significant. This theoretical perspective suggests systematic protocols for improving exponent extraction in other universality classes by adapting the detrending order.

Our comparative analysis with the Random Bond Ising Model (RBIM) demonstrates both the necessity and the discriminatory power of the proposed FSS protocol. At finite scales, disordered lattices exhibit a broad multifractal dispersion that partially originates from the same pre-asymptotic discreteness artifacts affecting the pure model — a finding that warns against interpreting single-snapshot results as intrinsic physics. However, the FSS analysis reveals a physically meaningful distinction: unlike the pure model, whose spectral width vanishes entirely in the thermodynamic limit, the RBIM retains a finite residual dispersion $\Delta H_{\infty} \approx 0.123$ that persists asymptotically. This residual is the genuine multifractal signature of Griffiths phases, detectable only after the transient artifacts have been filtered out by systematic scaling. Consequently, the FSS protocol is not merely a safeguard against false positives, but an active diagnostic tool: it simultaneously eliminates spurious finite-size broadening and isolates the asymptotic physical signal that single-scale analyses confound with noise..

The rigorous protocol established here enables MFDFA to function as a quantitative probe of universality classes in discrete lattice systems. By controlling finite-size effects through systematic FSS and choosing moment ranges to reflect the physical meaning of fluctuations, practitioners can reliably extract critical exponents and identify universality classes—even in experimental systems with unknown underlying Hamiltonians. The open-source implementation we provide ensures that this methodology is accessible to the broader computational physics community, fostering reproducible and reliable characterization of critical phenomena across disciplines and application domains.

Extensions of this framework offer promising directions for future research. Application to three-dimensional systems (3D Ising, $O(N)$ models) will quantitatively test our predictions for exponents in different universality classes. Adaptation to dynamical critical phenomena and quantum phase transitions will explore whether the RG-filter interpretation of detrending extends beyond equilibrium statistical mechanics. Systems with emergent scale invariance—including self-organized criticality and complex networks—represent another frontier where careful moment-range selection may illuminate genuine multifractal signatures. Most ambitiously, this work suggests a deeper connection between signal processing and RG theory: polynomial detrending, typically viewed as a technical necessity, may implement a systematic suppression of irrelevant operators, opening new avenues for extracting universal physics from finite-system measurements.

Beyond theoretical spin models, the methodological vulnerabilities and the FSS protocol established in this work have immediate and critical implications for the analysis of real-world empirical data. MFDFA is extensively applied across diverse physical domains—ranging from the analysis of ocean-atmosphere climate indices (e.g., the Tropical Southern Atlantic Index, TSA) to the study of chaotic rotational dynamics in astrophysics via digitalized light curves (e.g., Saturn's moon Hyperion).

It is crucial to recognize that empirical time series are fundamentally analogous to the discrete lattices studied here: they are strictly finite in length and subject to digital sampling (discretization). Our results demonstrate that applying MFDFA to a single, finite digitalized record inherently risks generating spuriously broad singularity spectra—often mistakenly interpreted as genuine multifractal complexity—due to the mathematical divergences associated with discrete step-like fluctuations at small scales. Therefore, applying an analog of our proposed scaling protocol (such as systematically analyzing subsets of varying total length N to test for the asymptotic collapse of the H(q) dispersion) is a necessary safeguard. It ensures that the reported multifractality in experimental data reflects true underlying nonlinear multiplicative processes rather than the ubiquitous artifacts of finite digital sampling.

\bibliographystyle{apsrev4-2} 
\bibliography{references}     

\end{document}